\documentstyle[preprint,aps,epsf]{revtex}
\begin{document}
\title{Quantum Measurement of a Single Spin using
Magnetic Resonance Force Microscopy}
 \author{G.P. Berman,$\!^1$ F. Borgonovi,$\!^{1,2}$ G. Chapline,$\!^{1,3}$ S.A. Gurvitz,$\!^{1,4}$ 
 P.C. Hammel$^5$,\\
 D.V. Pelekhov,$\!^5$ A. Suter,$\!^5$ and V.I. Tsifrinovich$^6$}
 \address{$^1$Theoretical Division and CNLS,
 Los Alamos National Laboratory, Los Alamos, NM 87545}
 \address{$^2$Dipartimento di Matematica e Fisica, Universit\`a Cattolica,
 via Musei 41 , 25121 Brescia, Italy, and I.N.F.M., Gruppo Collegato
 di Brescia, Italy, and I.N.F.N., sezione di Pavia , Italy}
 \address{$^3$Lawrence Livermore National Laboratory, Livermore, CA 94551}
 \address{$^4$Department of Particle Physics, Weizmann Institute of Sciences, 
Rehovot 76100, Israel}
 \address{$^5$Condensed Matter and Thermal Physics, Los Alamos National Laboratory, 
MS K764, Los Alamos NM 87545}
 \address{$^6$IDS Department, Polytechnic University,
 Six Metrotech Center, Brooklyn NY 11201}
\maketitle
{\bf Single-spin detection is one of the important challenges facing the development of several new technologies, 
e.g. single-spin transistors and solid-state quantum computation. Magnetic resonance force microscopy with a cyclic 
adiabatic inversion, which utilizes a cantilever oscillations driven by a single spin, is a
promising technique to solve this problem. 
We have studied the quantum dynamics of a single spin interacting with a quasiclassical cantilever.
It was found that in a similar fashion to the Stern-Gerlach interferometer the quantum dynamics generates a quantum superposition of two quasiclassical trajectories of the cantilever which are related to the two spin 
projections on the direction of the effective magnetic field in the rotating reference frame. 
Our results show that quantum jumps will not prevent a single-spin measurement
if the coupling between the cantilever vibrations and the spin is small in comparison with 
the amplitude of the radio-frequency external field.}\\ \ \\
Modern solid-state technologies are approaching the level at which manipulating a 
single electron, atom, electron or nuclear spin becomes extremely important. 
The future successful development of these technologies depends significantly on development 
of single-particle measurement methods. While 
a single electron charge can be detected using a single-electron transistor,  methods for detection 
of a single electron (or nuclear) spin in solids are still not available. However, many proposals for 
solid-state nano-devices require a single-electron (or nucleus) 
spin measurement. There are a 
few proposals for a solid-state single-spin measurement based on the
swap operation of a  spin state to a charge 
state \cite{kane}, scanning tunneling microscopy \cite{manassen},
or magnetic resonant force microscopy (MRFM)\cite{sidles1,sidles2}. 

In this paper, we consider a MRFM single-spin measurement. One of the most promising MRFM techniques 
is based on the cyclic adiabatic inversion (CAI) of electron or
nuclear spins \cite{rugar1}. In this technique,
the frequency of the spin inversion is in the resonance with the frequency of the mechanical vibrations of the 
ultrathin cantilever, which allows one to amplify the extremely weak force of a spin on the cantilever.  
The CIA MRFM method was successfully 
implemented as an alternative to the electron and nuclear magnetic resonance for macroscopic 
ensembles of spins \cite{rugar1,wago}. It has already achieved a sensitivity, which is equivalent to detection
of approximately 200 polarized electron spins\cite{bruland}. 

It is clear that the resonant amplification of the cantilever vibrations by a single spin
cannot be considered as a classical process. Indeed, the driving force acting on the cantilever 
is quantized, since it is proportional to the spin projection. As a result, 
quantum jumps could appear in the cantilever motion, which might prevent   
the resonant amplification of the cantilever oscillations. 
In fact, the problem of quantum jumps is a very general one. It 
always arises in a continuous observation of 
a single quantum particle. Despite extensive study\cite{knight}, 
the quantum jumps in a mechanical motion of classical detectors and their
effect on a measurement of a quantum system have not been investigated. The analysis 
of these quantum effects and their influence on a single-spin detection in CAI MRFM are the main subjects of 
this paper.

We consider the cantilever-spin system shown in Fig. 1. A single spin
($S=1/2$) is placed on the cantilever tip. The tip can oscillate only
in the $z$-direction. The ferromagnetic particle, whose magnetic
moment points in the positive $z$-direction, produces a non-uniform
magnetic field which acts on the spin. The uniform magnetic field, $\vec B_0$,
oriented in the positive $z$-direction, determines the ground state of
the spin. The rotating radio frequency ({\em rf}) magnetic field, $\vec B_1$, induces
transitions between the ground state and the excited state of the spin.
The origin is chosen to be the equilibrium position of the cantilever
tip with no ferromagnetic particle. The {\em rf} magnetic field can
be written as $B_x=B_1\cos [\omega t+\varphi(t)],~B_y=-B_1\sin [\omega t+\varphi(t)]$, 
where $\varphi(t)$ is a periodic change in phase with the frequency 
of the cantilever, required for a CAI of the spin. A non-uniform magnetic field produces a force on 
the spin which depends on the spin direction. If the spin direction is changed with a frequency 
which equals to the cantilever resonant frequency, the amplitude of the cantilever vibrations increases 
so that it can be detected by optical methods. 

In the reference frame rotating with the frequency
of the transversal magnetic field, $\omega+d\varphi /dt$, 
the Hamiltonian of
the ``cantilever-spin'' system is,
\begin{equation}
{\cal H}={{P^2_z}\over{2m^*_c}}+{{m^*_c\omega_c^2Z^2}\over{2}}
-\hbar\Bigg(\omega_L-\omega-{{d\varphi}\over{dt}}\Bigg)S_z
-\hbar\omega_1S_x-g\mu{{\partial B_z}\over{\partial Z}}ZS_z\ .
\label{a1}
\end{equation}
In Eq.~(\ref{a1}), $Z$ is the coordinate of the oscillator which describes
the dynamics of the quasi-classical cantilever tip; $P_z$ is its
momentum, $m^*_c$ and $\omega_c$ are the effective mass and the
frequency of the cantilever (the mass of the cantilever is: $m_c=4m_c^*$); $S_z$ and $S_x$ are the 
$z$- and the
$x$-components of the spin; $\omega_L=\gamma B_z$ (at $z$=0) is its Larmor frequency;
$\omega_1=\gamma B_1$ is the Rabi (or nutation) frequency; $\gamma=g\mu/\hbar$ is 
the gyromagnetic ratio of the spin; $g$ and $\mu$ are the g-factor and the nuclear magneton. 
(We consider for definiteness a nuclear spin, but the results can be applied also to an electron spin.) 

We assume that $\omega=\omega_L$, which means that the average 
frequency of the {\it rf} field, $\omega$, is equal to the Larmor frequency of the spin in the permanent 
magnetic field. Using the dimensionless variables: $\tau=\omega_ct$  
and $z=Z\sqrt{m_c^*\omega_c/\hbar}$ the spin-cantilever dynamics
is described in the rotating frame, by the following Heisenberg operator equation,
\begin{equation}
d^2z/d\tau^2+z=2\eta S_z\ ,~~~~
d{\vec{S}}/d\tau=[\vec{S}\times\vec{B}_{eff}],
\label{a2}
\end{equation}
where $\vec {B}_{eff}=(\epsilon,0, -d\varphi/d\tau+2\eta z)$ 
is the dimensionless effective magnetic field, $\epsilon=\omega_1/\omega_c$ and 
$\eta=g\mu(\partial B_z/\partial Z)/(2\sqrt{m_c^*\omega_c^3\hbar})$.
Thus, our model includes two dimensionless parameters, $\epsilon$ and
$\eta$. The first one is the dimensionless amplitude of the {\em rf} field,
and the second one describes the interaction of a single spin with a cantilever 
mechanical vibrations due to a non-uniform  
magnetic field produced by the ferromagnetic particle. Note that the term $d\varphi/d\tau$  in $\vec {B}_{eff}$
is caused by the phase modulation of the transversal magnetic field. 
The other part of the $z$-component of the effective magnetic field, $2\eta z$, describes a nonlinear 
effect -- a back reaction of the cantilever vibrations on the spin.

If the adiabatic conditions ($|d^2\varphi/d\tau^2| \ll\epsilon^2$) are
satisfied and the nonlinear effects are small in comparison to the effects of the {\it rf} field,
the average spin is ``captured'' by the effective magnetic field. More
precisely, it precesses around 
the effective magnetic field in such a way that the angle between the directions of the average spin and 
the effective magnetic field approximately remains constant.  
In CAI MRFM, the value of $d\varphi/d\tau$ changes periodically with
the period of the cantilever ($2\pi$ in our dimensionless
variables). As a results the effective magnetic field and the average
spin change their directions with the same period. This leads to a  
resonant excitation of the cantilever vibrations.

To test our model, we considered the classical limit of the macroscopic number of spins and 
the classical cantilever. Replacing the operators $S_x$ and $S_z$ 
in Eq.~(\ref{a2})
by the sums of operators over all spins in the sample and neglecting the quantum correlation effects, 
we obtain the classical equations of motion for the total average spin and for the cantilever. 
We solved these classical equations numerically for the parameters 
corresponding to the experiment with protons in ammonium nitrate\cite{rugar1}. 
To estimate the amplitude of stationary vibrations of the cantilever within the Hamiltonian approach, 
we consider the time, $\tau=Q_c$, where $Q_c$ is the quality factor of
the cantilever. We obtained for the amplitude of stationary vibrations of the cantilever 
$Z_{max}\approx 15$ nm, which is close to the experimental value in \cite{rugar1}, 
$Z_{max}\approx 16$ nm.

Since a cantilever can be considered as a quasi-classical measuring device, one might expect smooth 
increase of the cantilever amplitude also in a measurement of a single spin. 
However, a single spin z-component can accept only two values, $s_z=\pm 1/2$.
Therefore, smooth resonant vibrations of the quasi-classical
cantilever driven by the continuous oscillations of
$s_z$ seam to violate the principles of quantum mechanics. From this point of view, one should face 
quantum jumps rather than smooth oscillations of the cantilever. Such jumps could prevent the increase of 
the amplitude of the cantilever vibrations 
and a single-spin detection. To resolve this problem we consider the
quantum dynamics of the single spin-quasiclassical 
cantilever system.
The dimensionless Schr\"odinger equation can be written in the form,
\begin{equation}
i{{\partial\Psi(z,s_z,\tau)}\over{\partial\tau}}={\cal H}^\prime\Psi(z,s_z,\tau ),~~~~
\Psi(z,s_z,\tau)=\left(\matrix{\Psi_1(z,\tau)\cr
\Psi_2(z,\tau)\cr}\right)\ ,
\label{a5}
\end{equation}
where $\Psi(z,s_z,\tau)$ is a dimensionless spinor, ${\cal H}^\prime={\cal H}/\hbar\omega_c$ 
is the dimensionless Hamiltonian.  The functions
$\Psi_{1,2}(z,\tau)$ are the complex amplitudes to find a spin in the state $|s_z=\pm 1/2\rangle$ and a cantilever at the
point $z$ at time $\tau$.

To describe the cantilever as a sub-system close to the
classical limit, we choose the initial wave function of the
cantilever in the coherent state $|\alpha\rangle$. Namely, 
it was taken in the form, 
$ \Psi_1(z,0)=\sum_{n=0}^\infty A_n(0)|n\rangle,~\Psi_2(z,0)=0,~ $
and
$ A_n(0)=(\alpha^n/n!)\exp(-|\alpha|^2/2), $
where $|n\rangle$ is an eigenstate of the  oscillator
(cantilever) without spin. The initial average values of the coordinate and 
momentum are related to $\alpha$ as 
$ \langle z\rangle={{1}\over{\sqrt{2}}}(\alpha^*+\alpha),~\langle
p_z\rangle={{i}\over{\sqrt{2}}}(\alpha^*-\alpha)$. In order to correspond the 
classical limit, we took $|\alpha|^2\gg 1$.  

For numerical simulation of the single-spin-cantilever dynamics we used 
the value of the interaction parameter $\eta =0.3$. Currently this value is not feasible in 
experiments with a nuclear spin, but can be achieved in experiments with a single 
electron spin. For instance, for the cantilever parameters from\cite{stowe} and the magnetic field gradient
reported in\cite{bruland}, we obtain for a single electron spin 
$\eta =0.8$. The phase modulation of the {\em rf} field was taken in the form
$d\varphi/d\tau=-6000+300\tau$, (if $\tau \le 20$), and $d\varphi/d\tau=1000\sin(\tau-20)$,  
(if $\tau>20$), and the {\em rf} field amplitude, $\epsilon =400$. For these parameters 
the effective magnetic field,  $2\eta z$, produced by the cantilever vibrations on the spin 
remains small with respect to the amplitude of the {\em rf} field.

The numerical simulations of the quantum dynamics reveal the
formation of the asymmetric quasi-periodic Schr\"odinger cat (SC) state 
for the cantilever. 
Fig.~2 shows the typical probability distributions, $P(z,\tau )=|\Psi_1(z,\tau )|^2+|\Psi_2(z,\tau )|^2$, 
for nine instants of time. Near $\tau=40$, 
the probability distribution, $P(z,\tau)$, splits in two asymmetric peaks. After this the separation 
between these two peaks varies periodically in time. The ratio of the peak amplitudes is about 1000 
for chosen parameters (the probabilities are shown in the logarithmic scale).

It is clear that the small peak does not significantly influence the average coordinate of the cantilever. 
Fig.~3 shows the average 
coordinate of the cantilever, $\langle z(\tau)\rangle$, and the corresponding standard deviation,
$\Delta (\tau )=[\langle z^2\rangle -\langle z\rangle^2]^{1/2}$. 
One can see fast increase of the average amplitude of the cantilever
vibrations, while the standard deviation 
still remains small. This, in fact, is related to the initial
conditions of the spin, which was taken in the
direction of the $z$-axis. For instance, if the spin initially points
in the $x$-axis ($\Psi_1(z ,0)=\Psi_2(z,0)$),
our calculations show two large peaks with similar amplitudes.  

The two peaks in the cantilever probability distribution, shown in
Fig.~2, indicate two possible trajectories of the 
cantilever. As a result of the consequent measurement of the cantilever position the system
selects one of the two trajectories.     
The crucial problem is the following: Do the two peaks of the SC state
of the cantilever correspond to the definite 
spin states? To answer this question we studied the structure of the
wave function of the cantilever-spin system. 
First we have found that both functions, $\Psi_1(z,\tau)$ and $\Psi_2(z,\tau)$, contribute to each peak.
Fig. 4 illustrates the probability distributions, $|\Psi_1(z,\tau)|^2$
(red) and $|\Psi_2(z,\tau)|^2$ (blue),
for nine instants of time. One can see that these functions have maxima at the same
values of $z$. Next, we analyzed the structures of 
the functions, $\Psi_1(z,\tau)$ and $\Psi_2(z,\tau)$. When two peaks
are clearly separated we can represent each of these 
functions as a sum of two terms, corresponding to the ``big'' and ``small'' peaks,
\begin{equation}
\Psi_{1,2}(z,\tau)=\Psi_{1,2}^b(z,\tau)+\Psi_{1,2}^s(z,\tau).
\label{a6}
\end{equation}  
We have found that with accuracy up to 1\%
the ratio, $\Psi_1^s(z,\tau)/\Psi_2^s(z,\tau)=-\Psi_2^b(z,\tau)/\Psi_1^b(z,\tau)=\kappa (\tau )$,
where the $\kappa (\tau )$ is a real function independent of $z$. 
As a result, the total wave function can be represented in the form,
\begin{equation}
\Psi(z,s_z,\tau)=\Psi^b(z,\tau)\chi^b(s_z,\tau)+\Psi^s(z,\tau)\chi^s(s_z,\tau),
\label{a7}
\end{equation}
where $\chi^b(s_z,\tau)$ and $\chi^s(s_z,\tau)$ are spin wave functions, which are orthogonal to each other. 
Eq.~(\ref{a7}) shows 
that each peak in the probability distribution of the cantilever
coordinate corresponds to a definite spin wave function. 
We found that the average spin corresponding to the big peak
$\langle\chi^b|\vec S|\chi^b\rangle$
points in the direction of the vector $(\epsilon,0,-d\varphi/d\tau)$, whereas 
$\langle\chi^s|\vec S|\chi^s\rangle$ points in the opposite direction. 
Note that up to a small term, $2\eta z$, the vector 
$(\epsilon,0,-d\varphi/d\tau)$ is the effective magnetic field 
acting on the spin (see Eq.~(\ref{a2})).
The ratio of the integrated probabilities ($\int P(z,\tau )dz$)
for the small and big peaks ($\sim 10^{-3}$ in Fig. 2) 
can be easily estimated as $\tan^2(\Theta/2)$, where $\Theta$ is the
initial angle between the effective field, 
$(\epsilon,0,-d\varphi/d\tau)$, and the spin direction. Therefore by
measuring the cantilever vibrations, one finds
the spin in a definite state along or opposite to the effective magnetic
field. Our numerical simulations show that starting 
with such a new initial condition, i.e. when the average spin points along or 
opposite to the effective field, 
the probability distribution $P(z,\tau )$ shows again two peaks,
as in Fig.~2, but the ratio of the integrated probabilities of these peaks is much less ($\sim 10^{-6}$).
Thus for chosen parameters, the quantum jumps generated by a single spin measurement 
cannot prevent the amplification of the cantilever vibration amplitude, and thus the detection of 
a single spin. 

So far, the described picture reminds the classical Stern-Gerlach effect in which the cantilever measures 
the spin component along the effective magnetic field. An appearance of the second  peak, 
even if the average spin points initially in the direction of the effective magnetic field, 
provides a difference with the Stern-Gerlach effect. 
The origin of this peak is a small 
deviations from the adiabatic motion of the spin even at large amplitude of the effective field, and the back reaction of the cantilever vibrations on the spin. 
The next important question is the following: Is it possible to use
CAI MRFM to measure the state of a single spin? We 
studied numerically the phase of the cantilever vibrations when 
the initial spin  points along or opposite the direction of the effective magnetic field. Our computer simulations
show that the phases of the cantilever vibrations for these two
initial conditions are significantly different. 
When the amplitude of the cantilever vibrations increases, the phase difference for 
two initial conditions approaches $\pi$. Thus, the classical phase of the
cantilever vibrations indicates the state of the spin 
relatively to the effective magnetic field. If the spin is initially
in the superposition of these two states, 
it will acquire one of these states in the process of measurement.

In practical applications it would be very desirable to use CAI MRFM
for measurement of the initial $z$-component 
of the spin. 
For this purpose one should provide the initial direction of the
effective magnetic field to be a $z$-direction. Then, 
the initial $z$-component of the spin will coincide with its component
relatively to the effective magnetic field. 
In our computer simulations presented in Figs.~2-4 we have assumed
instantaneous increase of the amplitude of the {\it rf} 
field, at $\tau=0$. It causes the initial angle between the directions
of the spin and the effective magnetic field, 
$\Theta\approx\epsilon/|d\varphi/d\tau|\approx 0.07$. To eliminate
this initial angle we simulated the quantum 
spin-cantilever 
dynamics for adiabatic increase of the {\it rf} field amplitude:
$\epsilon=20\tau$ for $\tau\le 20$, and $\epsilon=400$ for 
$\tau>20$. Dependence for $d\varphi/d\tau$ was taken the same as in
Figs.~2-4. The results of these simulations are 
qualitatively similar to those presented in Figs.~ 2-4, but the
integrated probability of the small peak was reduced 
to its residual value $\sim 10^{-6}$. Neglecting this small
probability one can provide the measurement of the initial 
$z$-component of the spin. 

We should also mention that the detection of a single electron spin in an atom can be used to determine 
the state of its nuclear spin \cite{berman1,berman2}. Such a measurement is possible for an atom with a large 
hyperfine interaction in a high external magnetic field, because the electron spin frequency of the atom depends 
on the state of its nuclear spin.

In conclusion, 
we have analyzed the quantum effects in the single-spin measurement using cyclic adiabatic inversion (CAI) to drive cantilever vibrations in magnetic resonance force microscopy (MRFM).
We investigated the quasi-classical cantilever interacting with a single spin using Hamiltonian approach. 
We have shown that the spin-cantilever dynamics generates a Schr\"odinger-cat (SC) state for the cantilever. 
The two peaks of 
the probability distribution of the cantilever coordinate correspond
approximately to the directions of the spin along 
or opposite to the direction of the effective magnetic field, in the rotating frame. 

In this paper, we did not discuss the intriguing possibility of observing the SC state. This requires an 
estimate of the  SC life time, which cannot be derived using our Hamiltonian which neglects the environment of the cantilever. 
 Instead, we
concentrated on a possibility of observing 
the resonant excitation of the cantilever vibrations, driving by a
single spin. We demonstrated by a direct computation
of the average cantilever position and its standard deviation as a
function of time that the resonant amplification 
of the cantilever oscillations in indeed possible (for considered
region of the system parameters), despite the 
quantum jumps of the single spin. In fact, the standard deviation of
the cantilever coordinate becomes large only when the angle between 
the initial spin direction and the effective magnetic
field approaching $\pi /2$.  In this case the SC state appears with 
approximately equals peaks. However after an observation of the
cantilever position the system appears in one 
of the peaks, and the following evolution of the cantilever coordinate
shows again the resonant amplifications with 
a very small standard deviation. 

We expect that taking into consideration the interaction with an environment will not change our conclusion.
Such an interaction will cause the decoherence\cite{zurek}, which transforms the SC state
into the statistical mixture. It is clear that this effect as well as the thermal 
vibrations of the cantilever (see, for example\cite{rugar1,stowe}), cannot prevent an observation 
of the driven oscillations of the cantilever if the corresponding rms amplitude exceeds the rms amplitude of
the vibrational noise. Another effect of the interaction with the environment is the finite quality factor, $Q_c$, of the 
cantilever, which puts the limit on the increase of the cantilever vibrations. The stationary amplitude
of the cantilever vibrations can be estimated in our Hamiltonian approach by putting $\tau =Q_c$. 

Finally, we mention two other possible techniques for the cyclic spin inversion in MRFM. One of them is 
the standard Rabi technique. This assumes that in our notation $d\varphi/d\tau=0$, and $\epsilon=1$, i.e. the Rabi 
frequency equals to the cantilever frequency. This technique seems to be simpler that the CAI MRFM. But
the amplitude of the {\it rf} field, $\epsilon$, must be much greater than the effective field produced by 
the cantilever on the spin $2\eta z \ll\epsilon=1$. In this case, the force acting on the cantilever 
is very small, and the amplification of the driven cantilever vibrations requires a long time, 
i.e. a large cantilever quality factor.
Another technique assumes the application of short $\pi$-pulses which periodically change the direction of the spin in the 
time-interval, which is very short in comparison to the cantilever period \cite{berman3}. If the time interval 
between successive pulses equals half of the cantilever period, this 
technique  provides a resonant 
amplification of the cantilever vibrations.  Testing this technique in MRFM experiments is a challenging problem.
\section*{Acknowledgments}

We thank D.P. DiVincenzo, R.G. Clark, G.D. Doolen, H.S. Goan,
A.N. Korotkov, R.B. Laughlin, S. Lloyd, H.J. Mamin, G.J. Milburn, V. Privman, M.L. Roukes, D. Rugar, 
J.A. Sidles, K. Schwab  for useful discussions. This work  was
supported by the Department of Energy under contract
W-7405-ENG-36 and DOE Office of Basic Energy Sciences. The work of
GPB and VIT was partly supported by the National 
Security Agency (NSA) and by the Advanced Research and 
Development Activity (ARDA).

\newpage
\begin{figure}
\begin{center}
\epsfxsize 16cm \epsfbox{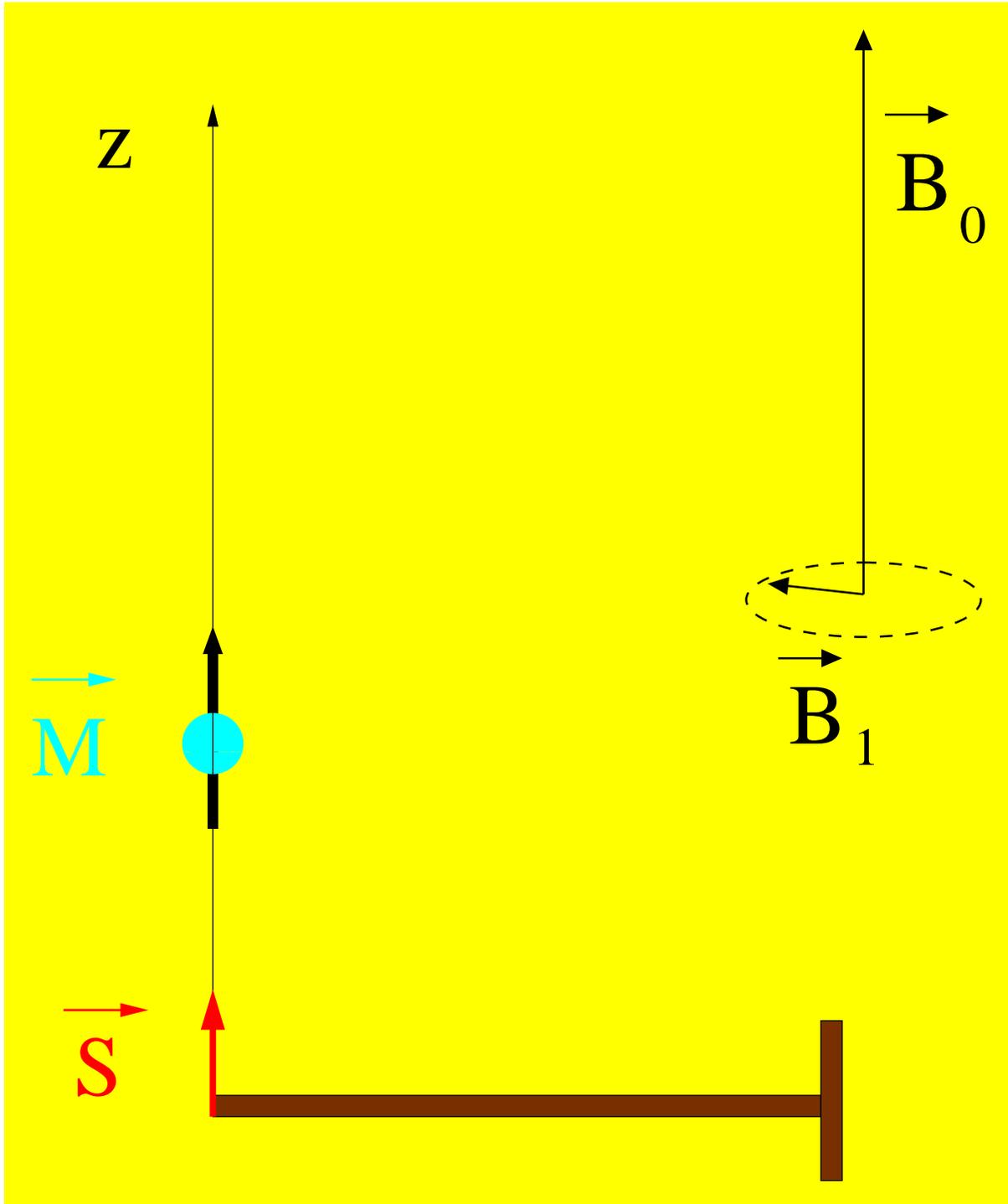} \narrowtext \caption{
The cantilever-spin system. $\vec{B}_0$ is the uniform
permanent magnetic field; $\vec{B}_1$ is the rotating magnetic field;
$\vec{ M}$ is the magnetic moment of the ferromagnetic particle; $\vec{S}$ is a single spin.}
\label{canti}
\end{center}
\end{figure}

\newpage
\begin{figure}
\begin{center}
\epsfxsize 16cm \epsfbox{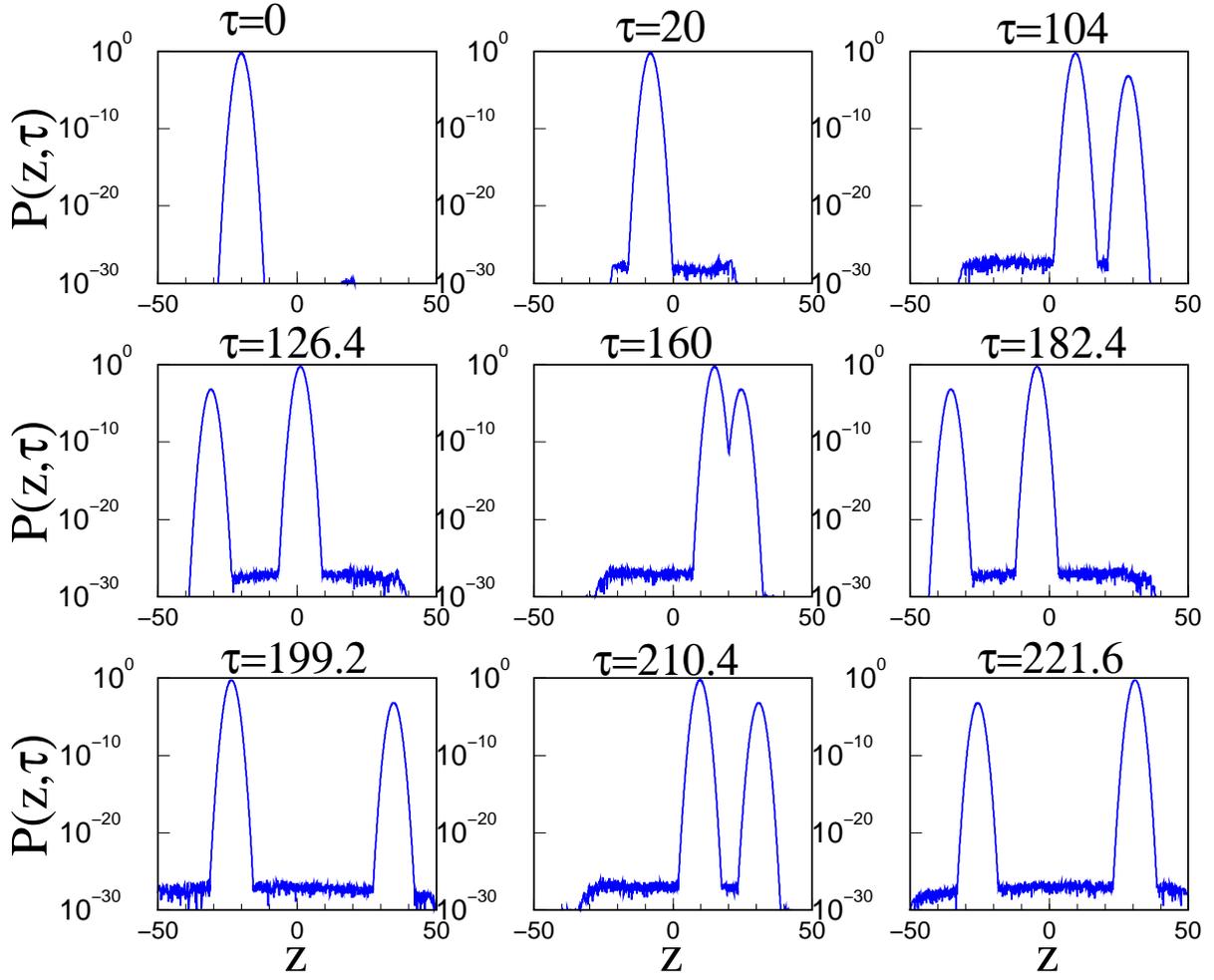} \narrowtext \caption{The probability
distribution, $P(z)$, for the cantilever position.
The values of parameters are: $\epsilon=400$
and $\eta=0.3$. The initial
conditions are: $\langle z(0)\rangle=-20$,
$\langle p_z(0)\rangle=0$
(which correspond to $\alpha=-10\sqrt{2}$).
The same parameters will be used in all figures below.}
\label{prob1}
\end{center}
\end{figure}

\newpage
\begin{figure}
\begin{center}
\epsfxsize 16cm \epsfbox{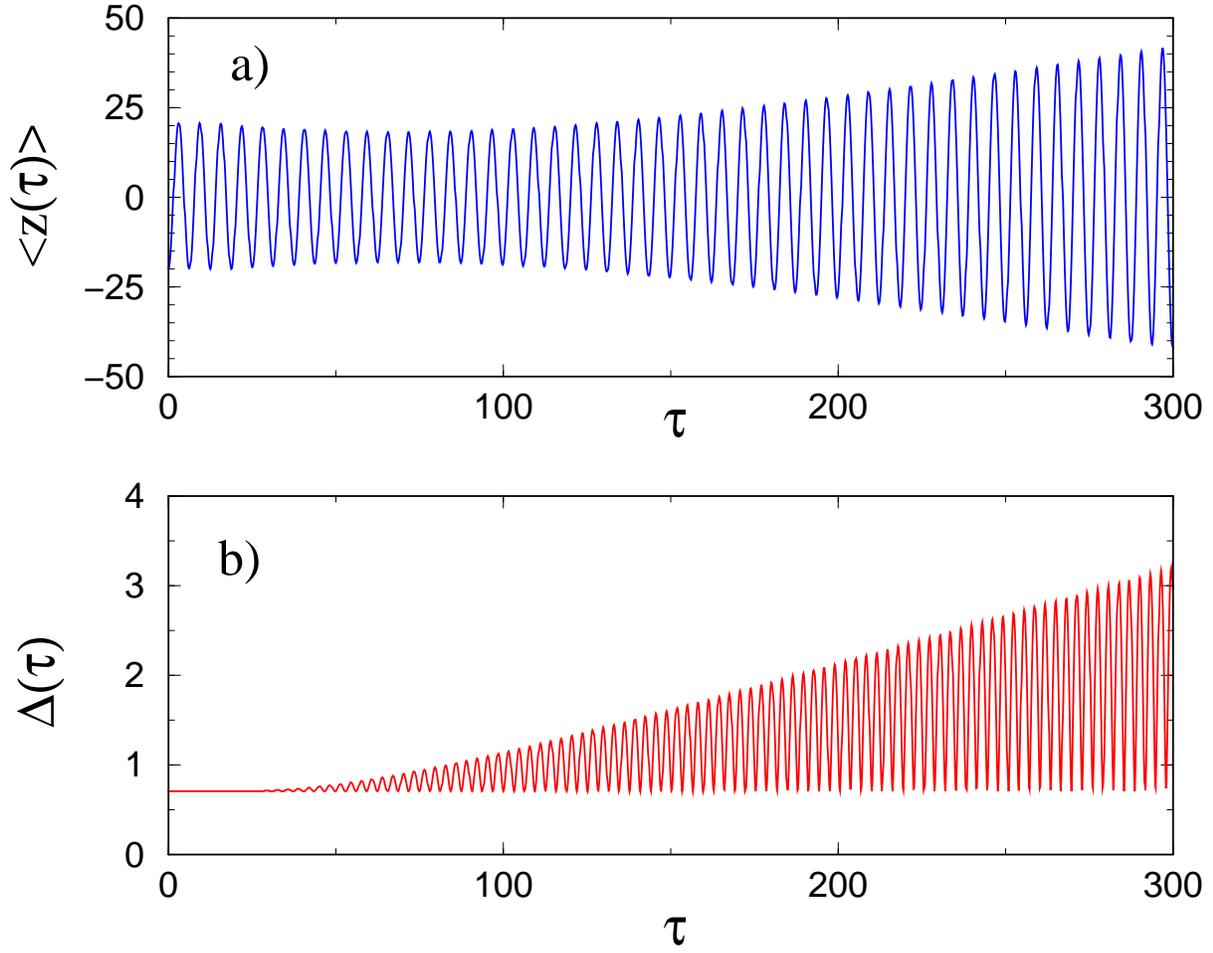} \narrowtext
\caption{Cantilever dynamics.
(a) Average coordinate of the cantilever as a function of $\tau$ and
(b) its standard deviation $\Delta (\tau) = [\langle z^2 (\tau) \rangle -
\langle z (\tau )\rangle^2]^{1/2}$.}
\label{ztau}
\end{center}
\end{figure}

\newpage
\begin{figure}
\begin{center}
\epsfxsize 16cm \epsfbox{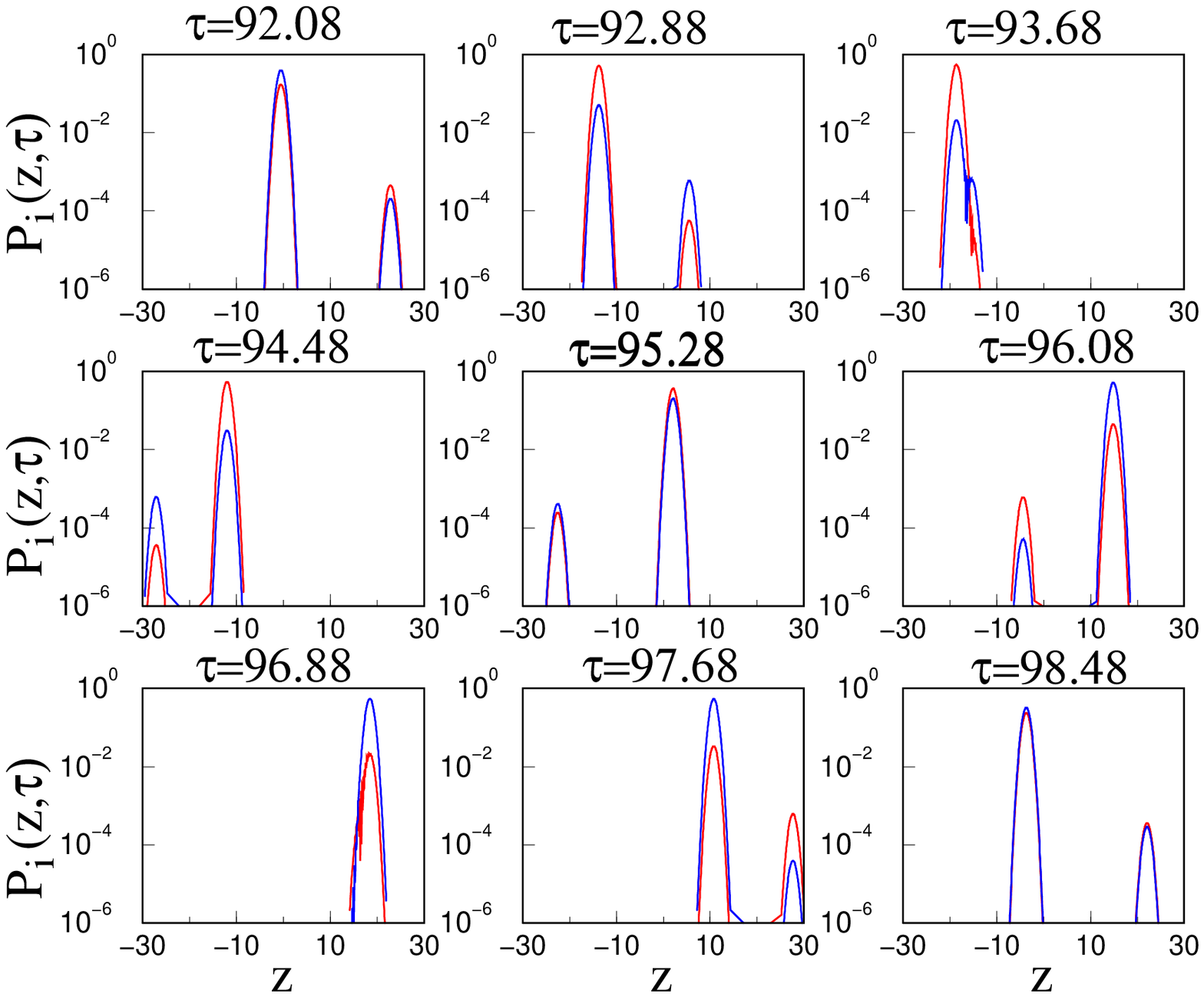} \narrowtext \caption{
The probability
distributions $P_i(z,\tau) =|\Psi_i(z,\tau )|^2$, 
$i=1$ (red) and $i=2$ (blue)
for nine instants in time: $\tau_k=92.08+0.8k$, $k=0,1,...,8$ }
\label{new4}
\end{center}
\end{figure}


\begin{thebibliography}{99}
%
\bibitem{kane} 
Kane, B.E. A silicon-based nuclear spin quantum computer. {\it Nature}. {\bf 393},  133-137  (1998).
%
\bibitem{manassen}
Manassen, Y., Mukhopadhyay, I. \& Rao, N.R. Electron-spin-resonance STM on iron atoms in silicon. 
{\it Phys. Rev. B}. {\bf 61}, 16223-16228 (2000).
%
\bibitem{sidles1}
Sidles, J.A. Nondestructive detection of single-proton magnetic-resonance.
 {\it Appl. Phys. Lett.} {\bf 58}, 2854-2856 (1991).
%
\bibitem{sidles2}
Sidles, J.A. Folded Stern-Gerlach experiment as a means for detecting nuclear-magnetic-resonance in individual nuclei. 
{\it Phys. Rev. Lett.} {\bf 68}, 1124-1127 (1992).
%
\bibitem{rugar1}
Rugar, D., Z\"uger, O., Hoen, S., Yannoni, C.S., Vieth, H.M. \& 
Kendrick, R.D. Force detection of nuclear-magnetic-resonance.  {\it Science}. {\bf 264}, 1560-1563  (1994).
%
\bibitem{wago}
Wago, K., Botkin, D., Yannoni, C.S. \& Rugar, D. Force-detected electron-spin resonance: Adiabatic inversion, nutation, 
and spin echo. {\it Phys. Rev. B}.
{\bf 57}, 1108-1114 (1998). 
%
\bibitem{bruland}
Bruland, K.J., Dougherty, W.M., Garbini, J.L., Sidles, J.A., Chao, S.H.
Force-detected magnetic resonance in a field gradient of 250 000 Tesla per meter.
{\it Appl. Phys. Lett.} {\bf 73}, 3159-3161 (1998).
%
\bibitem{knight} Plenio, M.B. and Knight, P.L.
The quantum jump approach to dissipative dynamics in quantum optics.
{\it Rev.Mod.Phys.} {\bf 70}, 101-144 (1998).
%
\bibitem{stowe}
Stowe, T.D, Yasumura, K, Kenny, T.W, Botkin, D, Wago, K. \& Rugar, D. Attonewton force detection using 
ultrathin silicon 
cantilevers. {\it Applied Phys. Lett.} {\bf 71} 288-290 (1997).
%
\bibitem{berman1}
Berman, G.P, Doolen, G.D, Hammel, P.C. \& Tsifrinovich, V.I. Solid-state nuclear-spin quantum computer 
based on magnetic 
resonance force microscopy. {\it
Phys. Rev. B}, {\bf 61}, 14694-14699 (2000).
%
\bibitem{berman2}
Berman, G.P, Doolen, G.D, Hammel, P.C. \& Tsifrinovich, V.I. Magnetic resonance force microscopy 
quantum computer with 
tellurium donors in silicon.  {\it Phys. Rev. Lett.}
{\bf 86}, 2894-2896 (2001). 
%
\bibitem{zurek}
Zurek, W.H. Decoherence and the transition from quantum to classical. {\it Physics Today}. {\bf 44}, 36-44 (1991).
%
\bibitem{berman3}
Berman, G.P. \& Tsifrinovich, V.I. Modified approach to single-spin detection using magnetic resonance 
force microscopy. 
{\it Phys. Rev. B.} {\bf 61}, 3524-3527 (2000).
%
\end{thebibliography}
\end{document}